\documentclass[aps,preprint,UTF8]{revtex4}
\usepackage{graphicx,amsmath,multirow,amssymb}
\draft
\begin{document}
\author{Bin Wang$^{a}$}
\address{$^{a}$ Department of Physics, Southwest Petroleum University, Nanchong 637001, P. R. China}
\title{System stability and truncation schemes to the Dyson-Schwinger Equations}
\begin{abstract}
With decades of years development, although important progresses have been made by the pioneers of this field, providing a sophisticated truncation scheme is still a great challenge up to now if the Dyson-Schwinger Equations(DSEs) of both quark and gluon propagators (or including even more DSEs) remain after truncation. In this work we view the coupled reminiscent DSEs of the gluon and quark propagators after truncation as a system with feedback. Then studying the stability of this equation array gives useful results.
Our calculation shows that the sum of the gluon and ghost loops plays the most important role in keeping this system stable and having reasonable solutions. The quark-gluon vertex plays a relative smaller but also important role. Our method also could give constraints and inspirations on fabricating a more sophisticated model of the quark-gluon vertex.

\bigskip


\bigskip
 E-mail:

\bigskip


\end{abstract}
\maketitle

\section{Introduction}
It is known that the relations of different points Green functions can be given by the Dyson-Schwinger Equations (DSEs) in nonperturbative field theories. Because any Green function in the DSEs is correlated to all other ones, it need be truncated in calculation\cite{roberts1994,roberts2000,roberts2003,MT1,MT2,QC,qc2,Fischer2003,Fischer2013}.
In the literature, the gap equation, i.e. the DSE of the quark propagator, can be solved with modeled quark-gluon vertex and gluon propagator. But there is great difficulty if both the quark and gluon DSEs (or including even more DSEs) remain after truncation.
Almost all truncation schemes up to now can not work in this case\cite{roberts1994,roberts2000,roberts2003,MT1,MT2,QC,qc2}, for example, the Maris-Tandy and Qin-Chang model had never been applied to coupled reminiscent DSEs beyond gap equation \cite{MT1,MT2,QC,qc2}. Although the truncation scheme given in \cite{Fischer2003,Fischer2013} can work well for the coupled reminiscent DSEs of the quark and gluon propagators, but it is still not perfect and need improve, for example it need use different quark-gluon vertexes in the quark self-energy and the quark loops respectively. What hinders people? We think that the lack of even crudely reasonable solutions to the coupled reminiscent DSEs in most of truncation schemes might hinder people mostly. Even worse, there are too many possible reasons to cause such difficulty since there are truncations and model given ingredients in the DSEs. At first of this work we will study this problem from the standpoint of system stability and will show that a very small suppression factor is needed to make truncation schemes work. Then we will show that the terms we dropped in truncation schemes will naturally provide such a small suppression factor. This procedure would give important inspirations at last.

At finite temperature and chemical potential the DSE of the quark propagator can be written as\cite{roberts1994,roberts2000,roberts2003}
\begin{equation}\label{gapeq}
G^{-1}(\widetilde{p_k})=i\gamma\cdot \widetilde{p_k}+m+\frac{4}{3}T\sum_{n=-\infty}^{+\infty}\int\frac{d^3q}{(2\pi)^3}g^2D_{\mu\nu}(\widetilde{p_k}-\widetilde{q_n})\gamma_{\mu}G(\widetilde{q_n})
\Gamma_{\nu}(\widetilde{q_n},\widetilde{p_k}).
\end{equation}
in which $\widetilde{p_k}=(\widetilde{\omega_k},\vec{p})=[(2k+1)\pi T+i\mu,\vec{p}]$,  $\widetilde{\omega_k}=(2k+1)\pi T+i\mu$, $m$ is the current quark mass (we use $m=5 \mathrm{MeV}$ in this paper), $G(\widetilde{p_k})$ is the full quark propagator, $D_{\mu\nu}(\widetilde{p_k}-\widetilde{q_n})$ is the full gluon propagator and $\Gamma_{\nu}(\widetilde{q_n},\widetilde{p_k})$ is the full quark-gluon vertex.
The inverse of the quark propagator $G^{-1}(\widetilde{p_k})$ can be decomposed as
\begin{equation}\label{eq2}
G^{-1}(\widetilde{p_k})=i\vec{\gamma}\cdot\vec{p}A(\widetilde{p_k}^2)+i\gamma_4\widetilde{\omega_k} C(\widetilde{p_k}^2)+B(\widetilde{p_k}^2).
\end{equation}
The  quark-gluon vertex $\Gamma_\nu$ is generally given by model, the gluon propagator $D_{\mu\nu}$ is also given \cite{MT1,MT2,QC,qc2} or partly given \cite{Fischer2003,Fischer2013} by model.
Several most popular truncation schemes in this field \cite{MT1,MT2,QC,qc2,Fischer2003,Fischer2013} can give very good value of hadron properties and the reasonable critical temperature.
One generally used truncation scheme is the bare approximation of the quark-gluon vertex, i.e. $\Gamma_\mu\rightarrow\gamma_\mu$, and the Qin-Chang model gluon propagator (in Landau gauge)\cite{QC}
\begin{equation}
[g^2D_{\mu\nu}(Q;\mu=0)]_{QC}=g^2\Delta(Q^2)(\delta_{\mu\nu}-\frac{Q_{\mu}Q_{\nu}}{Q^2})
=\frac{4\pi^2}{\omega^4}D_0e^{-{Q^2}/{\omega^2}}(\delta_{\mu\nu}-\frac{Q_{\mu}Q_{\nu}}{Q^2})\label{qc}
\end{equation}
in which $Q=(\vec{p}-\vec{q},\tilde{\omega_k}-\tilde{\omega_n})$, ${\omega}D_0 = (0.80\mathrm{GeV})^3$, $\omega=0.548$. With this truncation scheme and the chosen parameters the critical temperature of nuclear matter will be given as 150 MeV, which is consistent with Lattice QCD result \cite{Tc1,Tc2}.
\begin{figure}
  \includegraphics[width=12cm]{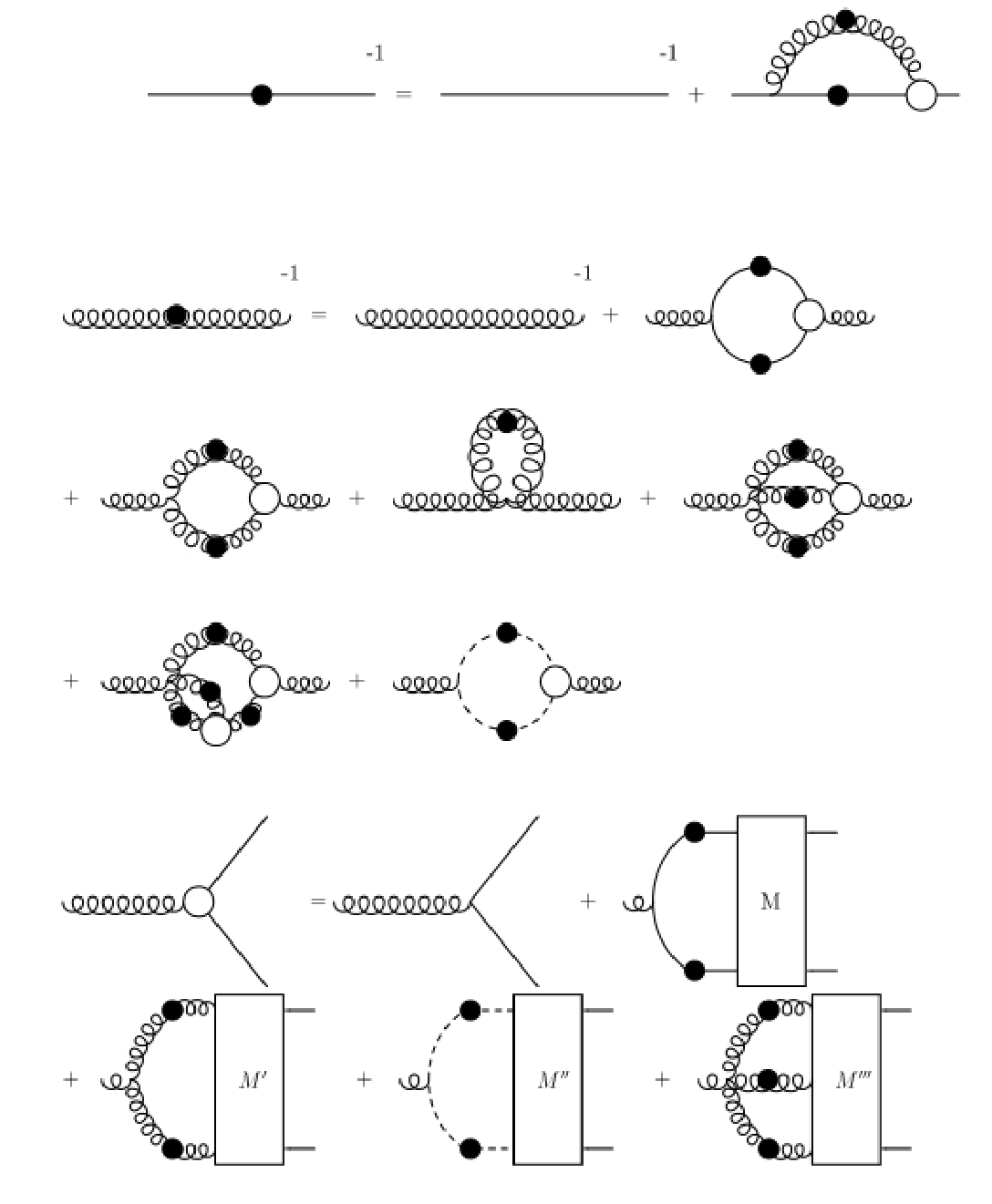}\\
  \caption{The DSEs for the quark propagator (top), the gluon propagator (middle) and the quark-gluon vertex(bottom).} \label{dse}
\end{figure}
\begin{figure}
  \includegraphics[width=12cm]{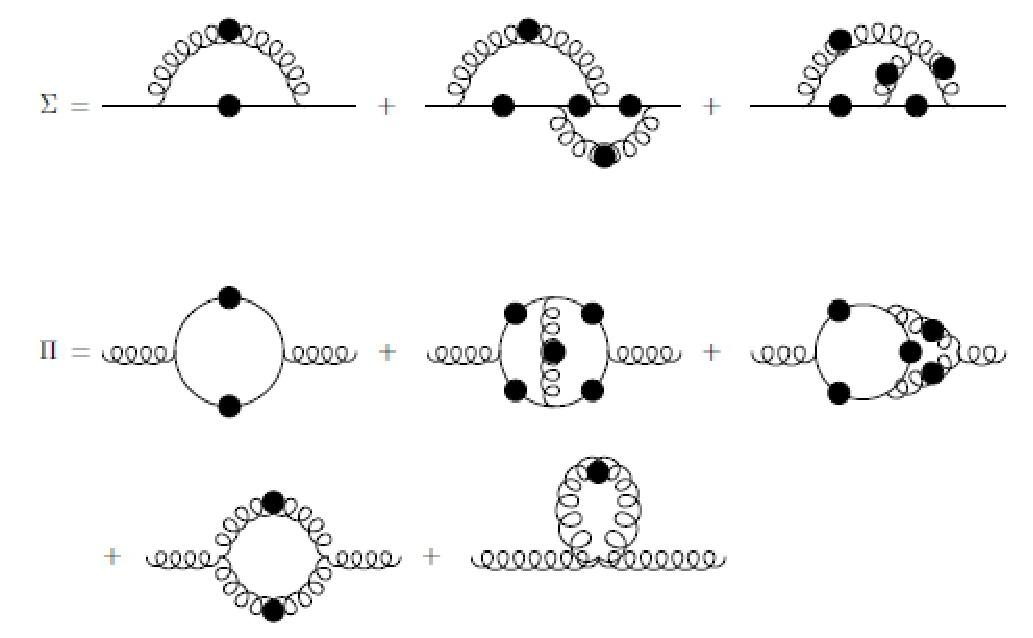}\\
  \caption{The lowest two orders of the quark and gluon self-energy (with respect to chemical potential dependent terms). We refer to the first diagram of $\Sigma$ and $\Pi$ as $\Sigma^{(1)}$ and $\Pi^{q(1)}$, the sum of second and third diagrams of $\Sigma$ as $\Sigma^{(2)}$, the sum of second and third diagrams of $\Pi$ as $\Pi^{q(2)}$,  the sum of fourth and fifth diagrams of $\Pi$ as $\Pi^{\mathrm{g(2)}}$.} \label{sigmapi}
\end{figure}

At zero chemical potential, as the previous truncation scheme gives reasonable results in calculating hadronic properties and critical temperature we can view it as physical one at first. When the DSE is extrapolated to finite chemical potential, the unquenching effect would make the gluon propagator change with quark chemical potential.
The baryon number chemical potential dependence of the gluon propagator can be included by gluon DSE. The difference between the inverse gluon propagators at nonzero and zero chemical potential can be written as
\begin{equation}\label{t1}
[g^{2}D_{\mu\nu}(Q;\mu)]^{-1}-[g^{2}D_{\mu\nu}(Q;0)]^{-1}=\Pi_{\mu\nu}(Q;\mu)-\Pi_{\mu\nu}(Q;0)\equiv\hat{\Pi}_{\mu\nu}(Q),
\end{equation}
in which $\Pi_{\mu\nu}(Q;\mu)$ represents the gluon self-energy divided by $g^2(Q^2)$. Since there is great difficulty in numerical calculation, only the quark loops contribution to the gluon self-energy can be included in practice. Then we can substitute this modified gluon propagator into the gap equation to calculate the quark propagator, but unfortunately no reasonable solutions exist, i.e. there is no stable solution or the solution at $\mu=1\mathrm{MeV}$ has much difference to the solution at $\mu=0$. It is known that for a system with feedback, it's stability will break if the feedback is strong enough. In this problem Eq.(\ref{t1}) can be viewed as a feedback to Eq.(\ref{gapeq}), and no reasonable solution existing shows the feedback is too strong and should be decreased.
Because $g^2$, $D_{\mu\nu}$ and $\Pi_{\mu\nu}$ are all model dependent in practice, the two sides of the previous equation should not be strictly equal and a correction factor $\eta(T,Q)$ need be added
\begin{equation}\label{t2}
[g^{2}D_{\mu\nu}(Q;\mu)]^{-1}_{QC}-[g^{2}D_{\mu\nu}(Q;0)]^{-1}_{QC}=\eta(T,Q)[\Pi^{QC}_{\mu\nu}(Q;\mu)-\Pi^{QC}_{\mu\nu}(Q;0)]\equiv\eta(T,Q)\hat{\Pi}^{QC}_{\mu\nu}(Q).
\end{equation}
The superscript/subscript $QC$ means depending on Qin-Chang model. In calculating the gluon self-energy $\Pi^{QC}_{\mu\nu}$, the gluon and ghost loops will be neglected and only the quark loops with bare vertex are included. The need of stability for the coupled reminiscent DSEs of gluon and quark propagators will confine the correction factor in a reasonable zone. At first we further neglect the momentum dependence of $\eta$, then we find that it must be very small which lies in $(\eta^{-}_\mathrm{up},\eta_\mathrm{up})$ (Table.\ref{tab1}). In next section we will show that such a small correction factor can be naturally provided by the diagrams deserted in the truncation scheme. (In fact in the truncation scheme given in \cite{Fischer2003,Fischer2013} the model quark-gluon vertex applied to quark loop can be considered as containing a effective suppression factor.)

(We find four properties about the correction factor $\eta$: (1)the imaginary part of the gluon self-energy needn't be suppressed and adding a big imaginary value to $\eta$ can hardly influence the stability; (2)if the gluon self-energy with n=0 (the four momentum is $Q=(2n\pi T,\vec{q})$) is excluded the $\eta_{up}$ will dramatically increase, for example at T=100 MeV it will increase from 0.073 to 0.39;  (3)Table.\ref{tab2} shows that the infrared and ultraviolet region needn't be greatly suppressed, so that the suppression is just a strong correlation effect; (4)if the correction factor makes the coupled reminiscent DSEs has reasonable solution at small $\mu$, it also works for big $\mu$.
According to these four properties, in the following we will take $\mu=1\mathrm{MeV}$ and only need concern the real part of $\eta(T,Q)$ with four momentum $Q=(0,\vec{q})$. We will take two $\vec{q}$s with $q^2=0.1,0.3$ for example.)

\begin{table}[htbp]
\centering \caption{Important physical quantities obtained with Qin-Chang model. $\eta_\mathrm{up}$ and $\eta^{-}_\mathrm{up}$ are momentum independent, they are the up and low borders of the correction factor defined in Eq.(\ref{t2}) which can make the equation array stable. The last six lines are the correction factors $\eta$ given by the method in section \uppercase\expandafter{\romannumeral2} at $Q=(0,\vec{q})$ with $q^2=0.1,0.3$. The subscript of $\eta$ denotes the value of $q^2$. The superscript of $\eta$ denotes the value of $\alpha_1$ and $\alpha_2$, in which $\mathrm{Re}[\eta]^{m}$ denotes the $\mathrm{Re}[\eta]$ with maximum module when the ratios $\alpha_1/\alpha_0$, $\alpha_2/\alpha_0$ and $\alpha_2/\alpha_2$ are synchronously confined in $\pm(1/5,5)$. The remains are the assistant quantities defined in section \uppercase\expandafter{\romannumeral2} which are calculated with $q^2=0.1$.}\label{tab1}
\begin{tabular}{|c|c|c|c|c|c|c|c|c|c|c|c|c|c|c|}
\hline
$T(\mathrm{MeV})$&80&100&120&140&160&180&200\\
\hline
$\eta_\mathrm{up}$&0.082& 0.073& 0.059& 0.047& 0.040& 0.038& 0.036\\
\hline
$\eta^{-}_\mathrm{up}$&-0.042 &-0.080 &-0.135 &-0.142 &-0.101 &-0.080 &-0.066 \\
 \hline
$\mathrm{Re}[\alpha_{0}]$&2.996  &  1.752  & -4.873  & -6.783  & -10.740 &  -7.583  & -7.992 \\
 \hline
$\mathrm{Im}[\alpha_{0}]$&-3.954 &  -1.817 &  -0.152  & -0.010 &   1.245  &  0.033  &  0.032 \\
 \hline
$\mathrm{Re}[n_0]$&2.942 &   2.146  &  1.228  &  1.079 &   0.879   & 1.002   & 0.976 \\
 \hline
$\mathrm{Im}[n_0]$&-0.329 &  -0.217 &   0.002 &  -0.001&    0.047   & 0.001   & 0.002 \\
 \hline
$\mathrm{Re}[n_1]$& -0.074 &  -0.072 &  -0.069 &   0.020&   -0.079 &  -0.074 &  -0.072 \\
 \hline
$\mathrm{Im}[n_1]$& -0.082 &  -0.116 &  -1.167  &  4.560  & -0.253  &  0.292  &  1.013 \\
 \hline
$\mathrm{Re}[n_2]$& -9.8 &    -9.7  &   -9.1  &   13.6   & -14.5  &  -14.3   & -14.8 \\
 \hline
$\mathrm{Im}[n_2]$& 0.129 &   0.003 &  -0.173  &  0.247  & -0.776  & -1.054  & -1.411 \\
 \hline
$\mathrm{Re}[\eta^{\alpha_1= \alpha_2=1}_{q^2=0.1}]$ & 0.020  &  0.021  &  0.021 &  -0.021   & 0.014  &  0.015   & 0.015\\
 \hline
$\mathrm{Re}[\eta^{\alpha_1= \alpha_2=\alpha_0 }_{q^2=0.1}]$ &0.001 &   0.005 &  -0.008   & 0.001 &  -0.003  & -0.003  & -0.002\\
 \hline
$\mathrm{Re}[\eta_{q^2=0.1}]^{m}$ &0.024  &  0.042  & -0.044  & -0.024  & -0.016 &  -0.016  & -0.013\\
 \hline
$\mathrm{Re}[\eta^{\alpha_1= \alpha_2=1}_{q^2=0.3}]$ &  0.021 &   0.020  &  0.021  & -0.017  &  0.013  &  0.014 &   0.014 \\
 \hline
$\mathrm{Re}[\eta^{\alpha_1= \alpha_2=\alpha_0 }_{q^2=0.3}]$ &  0.004  &  0.012  & -0.008 &  -0.002 &  -0.003 &  -0.003 &  -0.001\\
 \hline
$\mathrm{Re}[\eta_{q^2=0.3}]^{m}$ &-0.038  &  0.072  & -0.041  & -0.031 &  -0.014 &  -0.014 &  -0.012\\
 \hline
\end{tabular}
\end{table}

\begin{table}[htbp]
\centering \caption{The up limit of the correction factor while only the gluon self-energy $\Pi_{\mu\nu}(Q)$ with $Q^2_{low}<Q^2<Q^2_{up}$ is considered. The temperature is at 100MeV.}\label{tab2}
\begin{tabular}{|c|c|c|c|c|c|c|c|c|c|}
\hline
$(Q^2_{low},Q^2_{up})(\mathrm{GeV^2})$&(0,$\infty$)&(0,0.02)&(0,0.1)&(0.1,0.2)&(0.2,0.3)&(0.3,0.4)&(0.4,0.5)&(0.5,0.6)&(0.6,$\infty$)\\
\hline
$\eta_\mathrm{up}$&0.073& $>$10.0& 0.30&  0.20&  0.30&0.6&0.4&0.6& 0.75\\
 \hline
\end{tabular}
\end{table}

\begin{table}[htbp]
\centering \caption{Comparison of the Qin-Chang model (QC) \cite{QC} and the Lattice-QCD gluon propagator \cite{gluonlat}. $\beta$ is their ratio.}\label{tab3}
\begin{tabular}{|c|c|c|c|c|c|c|c|}
\hline
$Q^2(\mathrm{GeV^2})$&0.01&0.1&0.2&0.3&0.5&0.6&0.8\\
\hline
$\Delta(Q^2)^{Lat}$&155.51&139.06& 112.67&90.21 &55.99&45.43&28.43\\
 \hline
$\Delta(Q^2)^{QC}$&791.22&586.33 & 420.26 & 301.23&  154.76&110.93&56.99\\
 \hline
$\beta$&1/5.088&1/4.216&1/3.730&1/3.339&1/2.764&1/2.442&1/2.004\\
 \hline
\end{tabular}
\end{table}
\section{Origination of the small correction factor}
For simplicity we assume that the physical gluon propagator $g^{2}D_{\mu\nu}$ and the physical gluon self-energy $\Pi_{\mu\nu}$ can be obtained by multiplying scalar factors to $[g^{2}D_{\mu\nu}]^{QC}$ and $\Pi^{QC}_{\mu\nu}$ respectively,
\begin{equation}
 g^{2}D_{\mu\nu}(Q)={\beta}(T,Q) [g^2D_{\mu\nu}(Q)]^{QC},\label{eq6}
\end{equation}
\begin{equation}
\hat{\Pi}_{\mu\nu}(Q)=\beta_{\Pi}(T,Q)\hat{\Pi}^{QC}_{\mu\nu}(Q),\label{eq7}
\end{equation}
By substituting the previous two equations to Eq.(\ref{t1}) and comparing to Eq.(\ref{t2}) ,we would have
\begin{equation}
\eta={\beta}\beta_{\Pi}.\label{etabeta}
\end{equation}
According to the last section we only concern the real part of $\eta$
\begin{equation}\label{eq9}
\mathrm{Re}[\eta]={\beta}\mathrm{Re}[\beta_{\Pi}],
\end{equation}
because $\beta$ is real (see later).  (Strictly speaking, $\eta$, $\beta$ and $\beta_{\Pi}$ all should be tensors, but taking them as scalars is reasonable in this work since only the trace of corresponding equations are important. See the last paragraph of this section.)

If we use the Lattice-QCD gluon propagator to approximate the physical one, comparing it with the Qin-Chang model could give $\beta$, the results are shown in Table.\ref{tab3}.

Then we start to calculate $\beta_{\Pi}$. We name the third diagram of the gluon DSE (middle line of Fig.\ref{dse}) as $\Pi^{q}$ and name all following diagrams as $\Pi^{g}$, so the gluon self-energy $\Pi$ consists of $\Pi^{q}$ and $\Pi^{g}$. We define
\begin{equation}\label{eq10}
\hat{\Pi}=\Pi(\mu)-\Pi(0),\quad
\hat{\Pi}^q=\Pi^q(\mu)-\Pi^q(0),\quad
\hat{\Pi}^g=\Pi^g(\mu)-\Pi^g(0).
\end{equation}
$\hat{\Pi}^g$ must be a function of $\hat{\Pi}$, $\hat{\Pi}$ and $\hat{\Pi}^q$ are functions of the quark propagator, then if the quark propagator has a perturbation there would be response
\begin{equation}\label{eq11}
\Delta\hat{\Pi}=\Delta\hat{\Pi}^q+\Delta\hat{\Pi}^g(\hat{\Pi}).
\end{equation}
The lowest several orders of $\hat{\Pi}^q$ and $\hat{\Pi}^g$ are plotted in the second line of Fig.\ref{sigmapi}. According to the symbols given in the caption of Fig.\ref{sigmapi} we can write
\begin{equation}\label{eq12}
\Delta\hat{\Pi}^q=\Delta\hat{\Pi}^{q(1)}+\alpha_1\Delta\hat{\Pi}^{q(2)},\quad \Delta\hat{\Pi}^g(\hat{\Pi})=\alpha_2\Delta\hat{\Pi}^{g(2)}(\hat{\Pi}),
\end{equation}
in which $\hat{\Pi}^{q(1)}$ is of order $g^2$, $\hat{\Pi}^{q(2)}$ and $\hat{\Pi}^{g(2)}$ are of order $g^4$, $\alpha_1$ and $\alpha_2$ are the coefficients caused by neglecting higher order diagrams.
Substitute Eq.(12) to Eq.(11) and do trace, we can have
\begin{equation}\label{eq13}
\mathrm{tr}[\Delta\hat{\Pi}]=(1+\alpha_1 n_1)\mathrm{tr}[\Delta\hat{\Pi}^{q(1)}]+ \alpha_2 n_2 \mathrm{tr}[\Delta\hat{\Pi}]
\end{equation}
in which
\begin{equation}\label{eq14}
n_1=\frac{\Delta \mathrm{tr}[\hat{\Pi}^{q(2)}]}{\Delta \mathrm{tr}[\hat{\Pi}^{q(1)}]},\quad n_2=\frac{\Delta \mathrm{tr}[\hat{\Pi}^{g(2)}(\hat{\Pi})]}{\Delta \mathrm{tr}[\hat{\Pi}]}.
\end{equation}
In practice $n_2$ need be approximated by the following assumption for simplicity
\begin{equation}\label{eq15}
n_2\approx\frac{\Delta \mathrm{tr}[\hat{\Pi}^{g(2)}(\hat{\Pi}^{q(1)})]}{\Delta \mathrm{tr}[\hat{\Pi}^{q(1)}]}.
\end{equation}
Since  $\Pi^{QC}$ is $\Pi^{q(1)}$, according to Eq.(7) and (13), we can have
\begin{equation}\label{eq16}
\beta_\Pi=\frac{\mathrm{tr}[\hat{\Pi}]}{\mathrm{tr}[\hat{\Pi}^{q(1)}]}=\frac{1+\alpha_1 n_1}{1-\alpha_2 n_2}.
\end{equation}

$n_1$ and $n_2$ can be calculated by the Feynman diagrams in Fig.\ref{sigmapi}, the results are listed in Table.\ref{tab1}. Every ingredient of the Feynman diagrams in Fig.\ref{sigmapi} should take real physical value as much as possible. Here we choose the gluon propagator and the running coupling constant obtained by Lattice QCD \cite{gluonlat}. Because we think the previously truncated DSEs can give physically acceptable quark propagator, so we choose the quark propagator given by solving that coupled reminiscent DSEs (i.e. Eq.(\ref{gapeq}) and (\ref{t2})). The perturbation of the quark propagator is given by
\begin{equation}\label{eq17}
\Delta G=G_{\eta_1}-G_{\eta_2},
\end{equation}
in which $G_{\eta}$ is the quark propagator calculated by the coupled reminiscent DSEs with a correction factor $\eta$, here we choose $\eta_1=0.02, \eta_2=0.001$.

$\alpha_1$ and $\alpha_2$ can not be calculated directly and will be estimated by studying the quark self-energy. The DSE of quark propagator and the lowest two orders of the quark self-energy are given in Fig.\ref{dse} and Fig.\ref{sigmapi} respectively. Like the gluon self-energy, with the symbols given in the caption of Fig.\ref{sigmapi}, we can write
\begin{equation}\label{eq18}
\Delta\hat{\Sigma}=\Delta\hat{\Sigma}^{(1)}+\alpha_0\Delta\hat{\Sigma}^{(2)}
\end{equation}
in which $\hat{\Sigma}^{(1)}$ is of order $g^2$, $\hat{\Sigma}^{(2)}$ is of order $g^4$, $\alpha_0$ is the coefficient caused by truncating higher orders diagrams. The previous equation can be written as
\begin{equation}\label{eq19}
\alpha_0=\frac{n_{00}-1}{n_0},
\end{equation}
in which
\begin{equation}\label{eq20}
n_0=\frac{\Delta \mathrm{tr}[\hat{\Sigma}^{(2)}]}{\Delta \mathrm{tr}[\hat{\Sigma}^{(1)}]},\quad n_{00}=\frac{\Delta \mathrm{tr}[\hat{\Sigma}]}{\Delta \mathrm{tr}[\hat{\Sigma}^{(1)}]}.
\end{equation}
In previous equation $\Delta \mathrm{tr}[\hat{\Sigma}]/4$ is just the scalar part of the perturbation of quark propagator $\Delta G$, $\Delta\hat{\Sigma}^{(1)}$ and $\Delta\hat{\Sigma}^{(2)}$ can be calculated from the Feynman diagram in Fig.\ref{sigmapi}, then we can obtain $n_0$, $n_{00}$ and $\alpha_0$.
In Eq.(\ref{eq12}) and (\ref{eq18}) we want to represent the quantities $\Delta\hat{\Pi^{q}}$, $\Delta\hat{\Pi^{g}}$ and $\Delta\hat{\Sigma}$ by their Feynman diagrams of order $g^4$. We assume that those quantities can manifest their relative weights in their low order terms, for example, $\mathrm{tr}[\Delta\hat{\Pi}^{q(2)}]$ would be bigger (at least not much smaller) than  $\mathrm{tr}[\Delta\hat{\Pi}^{g(2)}]$ if $|\mathrm{tr}[\Delta\hat{\Pi}^{q}]|>|\mathrm{tr}[\Delta\hat{\Pi}^{g}]|$. With this assumption $\alpha_0$, $\alpha_1$ and $\alpha_2$ should be in the same scale.
Our calculation shows that the correction factor $\mathrm{Re}[\eta]$ will be always in $(\eta^{-}_{up},\eta_{up})$ when $\alpha_0$ is fixed and the ratios between $\alpha_0$,$\alpha_1$ and $\alpha_2$ lie in $\pm(1/5,5)$.
The $\mathrm{Re}[\eta]$ with maximum modulus under this condition ($\mathrm{Re}[\eta]^{m}$) is given in Table.\ref{tab1}, the special cases with $\alpha_1=\alpha_2=1$ and $\alpha_1=\alpha_2=\alpha_0$ are also listed there.  The results indicate that the smallness of the correction factors is robust to the value of $\alpha_1$ and $\alpha_2$. Now let us explore the reason for such robustness. The $n_{00}$ given in Eq.(\ref{eq20}) represents the deformation caused by bare vertex approximation, such deformation need be compensated by magnifying the model gluon propagator, so that $n_{00}$  can be used to estimate the $\beta$ defined in Eq.(\ref{eq6})
\begin{equation}\label{eq21}
\beta=\frac{1}{n_{00}}=\frac{1}{1+\alpha_0 n_0}.
\end{equation}
Take the special case $\alpha_0=\alpha_1=\alpha_2$ for example, there is
\begin{equation}\label{eq22}
\eta=\beta \beta_{\Pi}=\frac{1+\alpha_0 n_1}{1+\alpha_0 n_0}\frac{1}{1-\alpha_0 n_2}.
\end{equation}
The first fraction is the contribution of the full quark-gluon vertex, the second fraction is the contribution of the sum of gluon and ghost loops.
Look at Table.\ref{tab1}, $\mathrm{Re}[-n_2]$ is a big value, so the second fraction tends to cause great suppression. Table.\ref{tab1} also shows that $\mathrm{Re}[n_0]\gg\mathrm{Im}[n_0]$ and $\mathrm{Re}[n_0]\gg\mathrm{Re}[n_1]$, which can make the real part of the first fraction small and will be useful to keep Re[$\eta$] small (since in most cases the imaginary part of the second fraction is small). So that the properties of $n_0$, $n_1$ and $n_2$ together help $Re[\eta]$ keep small under different $\alpha_1$ and $\alpha_2$.

(Strictly speaking $\eta,\beta,\beta_{\Pi}$ and $\alpha_{0,1,2}$ all should be tensors, but it is also reasonable to take scalars here. In the quark propagator Eq.(\ref{eq2}) the scalar term $B$ is most easily affected by the gluon self-energy and most relevant to stability. Because $[B(\mu)-B(0)]$ is proportional to $\mathrm{tr}[\hat{\Sigma}]$ and approximately proportional to $\mathrm{tr}[\hat{\Pi}]$, it is $\mathrm{tr}[\hat{\Sigma}]$ and $\mathrm{tr}[\hat{\Pi}]$ that affect the stability. So that it is reasonable to take $\beta,\beta_{\Pi},\alpha_{0,1,2}$ as scalars, then $\eta$ also should be a scalar.)
\section{Conclusions}
At the beginning of this work we show that the coupled reminiscent DSEs of the quark and gluon propagator after truncation can not keep stable with simple truncation schemes. This is because the truncation could deform every ingredient of the DSEs. So that a correction factor should be introduced which is confined in a small zone by the need of system stability(see Eq.(\ref{t2})). By studying the response of the quark and gluon self-energy to the perturbations of the quark propagators, such small correction factors could naturally present. From Eq.(\ref{eq22}) we can see that two points
are important.
The first point is that comparing to the bare (quark-gluon) vertex the full vertex could effectively diminish the quark loop's contribution. Table.\ref{tab1} shows that $\mathrm{Re}[n_0]$ is at least several times of $\mathrm{Re}[n_1]$ and $\mathrm{Im}[n_0]$, this indicates that the dependence of $\mathrm{Re}[\hat{\Sigma}]$ ($\Sigma$ denotes quark self-energy) on the higher order terms of the quark-gluon vertex is much stronger than that of $\mathrm{Im}[\hat{\Sigma}]$ and $\mathrm{Re}[\hat{\Pi}^q]$ ($\Pi$ denotes quark loop).
Substitute them into the first fraction of Eq.(\ref{eq22}) would suppress the real part of the correction factors. In the literature the longitudinal structure of the vertex can be guided by considering the gauge covariance \cite{bc1,bc2}, the transverse structure can be partly determined by considering the requirements of multiplicative renormalisability\cite{cp1,cp2,cp3}, besides that, there is no other guides to exploring the remaining terms of the transverse structure except checking whether the outcomes with the vertex model are physical reasonable\cite{nat}. The properties provided by our method might be useful in determining the full vertex lorentz structure.
The second point is that the sum of the gluon and ghost loops $\hat{\Pi}^g$ can greatly suppress the contribution of the quark loop $\hat{\Pi}^q$ to the gluon self-energy $\hat{\Pi}$ (corresponding to the bigness of $\mathrm{Re}[-n_2]$).

The outcomes of this work are helpful in dispelling the fogs hindering people and would be useful in finding more sophisticated truncation schemes in the future.

\acknowledgments
Thanks for the support of the Nanchong Shi Xiao Ke Ji He Zuo Xiang Mu(NC17YS4014).

\end{document}